# More green space is related to less antidepressant prescription rates in the Netherlands: A Bayesian geoadditive quantile regression approach

Marco Helbich, Nadja Klein, Hannah Roberts, Paulien Hagedoorn, Peter P. Groenewegen

**Abstract**

*Background:* Exposure to green space seems to be beneficial for self-reported mental health. In this study we used an objective health indicator, namely antidepressant prescription rates. Current studies rely exclusively upon mean regression models assuming linear associations. It is, however, plausible that the presence of green space is non-linearly related with different quantiles of the outcome antidepressant prescription rates. These restrictions may contribute to inconsistent findings.

*Objective:* Our aim was: a) to assess antidepressant prescription rates in relation to green space, and b) to analyze how the relationship varies non-linearly across different quantiles of antidepressant prescription rates.

*Methods:* We used cross-sectional data for the year 2014 at a municipality level in the Netherlands. Ecological Bayesian geoadditive quantile regressions were fitted for the 15%, 50%, and 85% quantiles to estimate green space–prescription rate correlations, controlling for physical activity levels, socio-demographics, urbanicity, etc.

*Results:* The results suggested that green space was overall inversely and non-linearly associated with antidepressant prescription rates. More important, the associations differed across the quantiles, although the variation was modest. Significant non-linearities were apparent: The associations were slightly positive in the lower quantile and strongly negative in the upper one.

*Conclusion:* Our findings imply that an increased availability of green space within a municipality may contribute to a reduction in the number of antidepressant prescriptions dispensed. Green space is thus a central health and community asset, whilst a minimum level of 28% needs to be established for health gains. The highest effectiveness occurred at a municipality surface percentage higher than 79%. This inverse dose-dependent relation has important implications for setting future community-level health and planning policies.

**Keywords**

Mental health; antidepressants; exposures; green space; quantile regression; spatial epidemiology

**Graphical abstract**

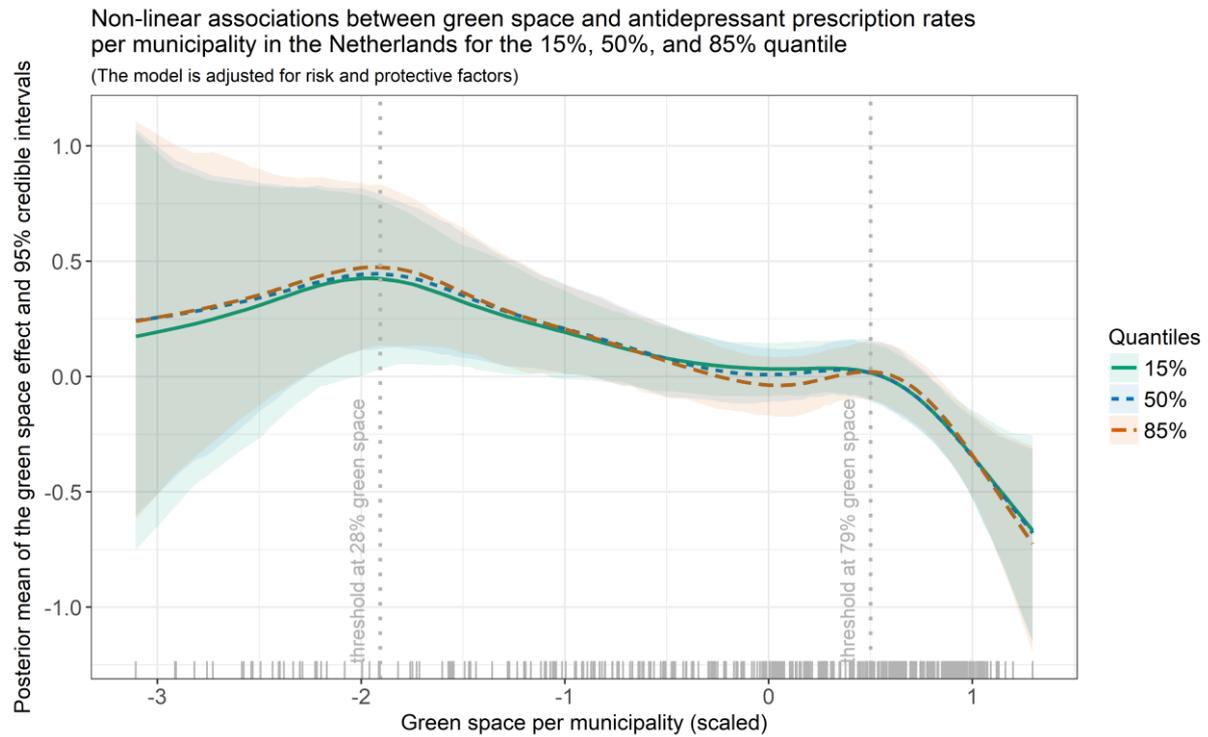

**Highlights**

- Green space was inversely correlated with antidepressant prescription rates.
- Bayesian geoadditive quantile regression showed non-linear dose-response functions.
- The shape of the associations showed moderate variations across quantiles.
- For health gains, communities should have at least one quarter green space; the most health gains occur when the proportion exceeds three quarters.

# 1. Introduction

Interest among both researchers and health policymakers in the mental health pathway of environmental exposure has grown substantially [1–5]. One reason for this is that neuropsychiatric conditions, such as depression, are now among the leading disease burdens globally [6,7]. This is of particular concern in the Netherlands, where the lifetime prevalence of depression is high (19%) [8].

Demographic factors, lifestyles, and household characteristics explain some, but not all, variation in depression prevalence [9–12]. There is much debate about the contribution of natural environments, such as green space (e.g., parks and woodland), within people's residential environment [3]. Findings that the presence of green space is among the determinants of people's mental health converge [4,13,14]. Although not consistently confirmed across experimental and epidemiological research, several cross-sectional [15–21] and a few longitudinal studies [22–25] show that green space exposures protect against the development and onset of poor mental health [26].

Because the mechanisms between green space and depression appear to be complex, the underlying pathways are under debate. Proposed interrelated mechanisms include, but are not limited to, green space restoring attention, supporting stress recovery and physical activity, and stimulating social interaction [13,14,26,27]. Empirical studies [15,17,28] testing these mechanisms typically rely on multiple-choice self-reported rating scales, instead of diagnostic interviews, to examine depression symptoms. This might provoke self-reporting response bias [29] and challenge comparability across studies through the availability of numerous depression screeners. This is accompanied by a lack of coherence in cut-off points to distinguish between mild, moderate, and severe depression [30].

These issues might be addressed by the use of objective health indicators such as medication prescription rates [31]. Antidepressant medication is widely utilized for depression treatment [32,33] and represents an ideal proxy for depression prevalence [34]. Apart from two contradictory studies in the UK [35,36], little is known about how green space is related to antidepressant prescription rates. However, such associative studies are more complicated than they seem at first sight. Linear mean regressions are fitted as standard methodology [15,18,22,28,36]. Lacking theoretical foundation [14,26], there is no plausible reason for such a simplification linking green space only to the conditional mean of the response variable (e.g., antidepressant prescription rates), which may over- or underestimate or incorrectly assume that no correlation exists [37]. To our knowledge, there is no research on how green space affects mental health for points other than the mean of the response distribution. Yet, it is rational to assume that the upper, central, and lower quantiles of the response variable may be affected differently by green space and may depend on different risk and protective factors. Bayesian geoadditive quantile regression (BQR) [38,39], which had not previously been applied in green space research, could provide valuable insights into other distributional features of the response that could not be uncovered with mean regressions. BQR also has the ability to account for non-linear effects on the quantiles and, importantly, is able to account for unobserved spatial heterogeneity. Non-linear dose-response functions [40] make intuitive sense because, for example, people living in areas with a high prevalence of depression (as indicated by a high prescription rate) may have different risk factors and exposures, and may benefit more from green space than those residing in healthier areas with lower prescription rates. That the functional form of green space is more complicated than linear was shown [41]. As area-level prescription data are spatially patterned [31,42], unexplained spatial heterogeneity needs to be incorporated into BQR, otherwise regression estimates will be biased [43].

The paucity of antidepressant rate–green space studies coupled with methodological concerns indicated a research need. We made a contribution to existing research by investigating the associations between the amount of green space and antidepressant prescription rates per municipality in the Netherlands. In order to pay particular attention to how the associations might vary across different quantiles, we applied BQR for the first time. We generated the following hypotheses: 1) The more green space in a municipality, the lower the antidepressant prescription rate, and 2) the strength of the association differs across the quantiles in a non-linear fashion. A more thorough understanding of this association may contribute to lower pharmaceutical spending on treating mental and behavioral disorders.

## 2. Materials and methods

### 2.1 Study background

As the Netherlands is among the top spenders on mental healthcare in the Organization for Economic Cooperation and Development (OECD) [44,45], we undertook the study in this country by means of a cross-sectional, ecological research design. The study was conducted at the municipality level to comply with privacy regulations and because this was the most detailed level where all data were accessible. We selected all 403 municipalities for 2014. Municipalities varied not only in size (median=6,495 ha; min.=696; max.=50,569), but also in population size (median= 25,691 people; min.=942; max=810,937).

### 2.2 Data

*Antidepressant prescription rates*

Depression prevalence, our outcome variable, was operationalized by means of antidepressant prescription rates per 1,000 inhabitants per municipality in the year 2014 [46]. Our data represent antidepressant medication prescribed by general practitioners (GPs), who are the first point of contact to treat symptoms of depression in the Dutch healthcare system. The antidepressants (N06A) are classified according to the Anatomical Therapeutic Chemical Classification System codes that GPs record per patient contact and include, for example, Selective Serotonin Reuptake Inhibitors such as Prozac. The antidepressant records were extracted from the Demand Supply Analysis Monitor (Vraag Aanbod Analyse Monitor) from the Primary Care Database 2014 [47], maintained by the Netherlands Institute for Health Services Research. The database comprised 391 GP practices with 1,541,902 listed patients. Municipalities with missing data on antidepressant prescriptions (N=19) were excluded.

*Green space*

The primary exposure variable of interest was green space. Given our ecological research design and the numerous ways to define green space [41], we defined green space as the proportion of green space per municipality (in %). Data on green space were gathered from the most recent Dutch land use database for the year 2012/13 [48]. We extracted and aggregated land use classes referring to parks, agricultural areas, forests, etc. using a geographic information system [16,49].

*Confounders*

For each municipality we collected nine confounders, informed by previous research. The data were uniformly aggregated to the municipality level. Unless stated otherwise, data comprised routinely collected information for 2014 obtained from Statistics Netherlands.

*Demographics*: Other studies have reported differences in mental disorders between the native population and migrants [50]. We therefore controlled for the proportion of non-Western people per municipality (in %). Because depression risk also varies across age cohorts, and older adults are at greater risk [51], we adjusted for the proportion of elderly people (in %).

*Socioeconomic status*: The absence of labor market participation may diminish a person's social status and increase the risk for depression [52]. We considered the unemployment rate (in %) among those aged 15–75 years. Neighborhood deprivation has also been shown to be correlated with depression [9]. In order to adjust for area-based deprivation, we included the average housing value (in 1,000 euros). We assumed that deprived areas are at higher risk and have noticeably higher antidepressant prescription rates.

*Health status and healthcare*: Physical activity protects against depressive disorders [53]. We used data on walking and cycling from a representative sample of approximately 71,000 participants aged 20–89 years in the Dutch National Travel Survey [54]. Physical activity encompasses the total average number of minutes spent walking and cycling per person per day per municipality. To obtain comprehensive estimates of physical activity levels, all trip purposes (e.g., leisure, commuting) were aggregated for 2010–14 while ensuring a sufficiently large number of respondents per area. As depression is often associated with higher mortality rates [55], we included an indirectly standardized mortality ratio[1] considering all causes of death to express an increased or decreased mortality risk at the population level. General practitioners (GPs) serve as gatekeepers for mental health treatment [47]. A better GP supply was expected to reduce depression prevalence. We modeled GP accessibility through the street network distance from each municipality to the closest GP [49].

*Urbanicity*: To adjust for urban–rural differences in psychiatric disorders [56], we included address density, namely the number of address locations per ha, abstracted from the cadaster (Basisregistraties Adressen en Gebouwen). Finally, a regional dummy variable, developed by the OECD, classifying each areas as either urban or rural, was used for comparative purposes.

## 2.3   Statistical analyses

Following descriptive summary statistics for each variable, we related low (≤15% quantile), mid-level (> 15% and <85% quantile), and high (≥85% quantile) antidepressant prescription rates to the corresponding green space per municipality. Non-parametric Wilcoxon rank sum tests were used to investigate differences across urban and rural areas statistically. Spearman's rank correlation coefficients $\rho$ were used to assess similarities between the variables.

Regression analyses were carried out in a Bayesian framework, which provided high flexibility in setting up models with varying complexity and allowed for non-linear associations of green space across municipalities. The BQR [39] model reads as

$$y_i = x_i'\beta_\tau + f_\tau(greenspace_i) + f_{geo,\tau}(municipalty_i) + \varepsilon_{i,t}, \quad i = 1, \dots, n,$$

and enables the analysis of the covariate effects semi-parametrically for each quantile separately. Above, $\tau \in (0,1)$ is the quantile of interest and $\varepsilon_{i,t}$ is the error term with cumulative distribution function $F_{\varepsilon_{i,t}}$. It is assumed that $\varepsilon_{i,t}$ and $\varepsilon_{j,t}$ are independent for $i \neq j$ and that the $\tau - th$ quantile of the error term distribution conditional on $x$ is zero, i.e. $F_{\varepsilon_{i,t}}(0|x) = \tau$. Furthermore, $x_i$ contains $p$ linear effects as well as a 1 for the overall intercept, $\beta_\tau$ is the $p+1$ dimensional vector of quantile specific regression coefficients for which a flat prior distribution is chosen, and $f_\tau(greenspace_i)$ was a non-linear smooth effect of green space that was modelled with Bayesian (P)enalized-splines [57]. We followed the literature and employed 22 cubic B-splines as basis functions on an equally spaced grid of knots to ensure a reasonably smooth but sufficiently flexible estimate of the non-linear effect. The coefficients were assigned second-order random walk priors. The variances of the random walk were estimated as well in a full Bayesian approach and we used a conjugate inverse gamma prior with small scale and shape parameter $a = b = 0.001$, which reflects a weakly informative prior. The assumption that the municipality effects were correlated and varied smoothly across neighboring municipalities was defined implicitly by a Gaussian Markov random field prior [58] and was captured through the effect $f_{geo,\tau}(municipalty_i)$. Two municipalities are neighbors if they share common borders. Such spatial correlation served as surrogate for covariates not included in the model. For an in-depth methodological discussion, we refer to [39].

---

[1] An SMR higher than 1 refers to more observed deaths than expected, while a value lower than 1 means fewer deaths than expected. When the observed number of deaths equals the expected one, the SMR is close to 0.

Our modeling strategy comprised two BQR models, each modeling the 15%, 50%, and 85% quantiles. Model 1 included green space in a linear fashion and the confounders. Model 2 extended the first, incorporating green space non-linearly. Both models acknowledged that adjacent municipalities might be spatially correlated. All variables were standardized on the same scale. To ensure convergence of the MCMC chains, we used a total of 52,000 iterations and discarded the first 2,000 iterations as burn-in. Further, to reduce autocorrelations, we stored each 50[th] sample such that our results were based on a final effective size of 1,000 (close to independent) samples of the posterior distribution.

For model comparison, we utilized the deviance information criterion (DIC). Lower DIC scores indicate a better model fit. For each model, we report summary statistics of the posterior distributions, including the mean together with the 95% credible intervals (CIs). To reach significance, 0 should not be contained in the CI. Statistical inference for the BQR models was carried out in BayesX 3.0.2 [59] and the R environment [60]. The BayesX scripts are provided in the supplementary materials.

## 3 Results

The median prescription rate per 1,000 persons per area was 330.5 (SD=50.0), ranging from 177.0 to 488.0. With a median of 288.0 units (SD=45.8) and 341.0 (SD=46.3) units, prescription rates varied significantly between urban and rural areas, as shown by the Wilcoxon test ($p<0.001$). The median of green space was 75.1% (SD=21.1%). Rural areas obviously have more green space (median=79.4%, SD=18.6%) than urban municipalities (median=62.7%, SD=24.0%). The difference is statistically significant ($p<0.001$). Table 1 provides additional descriptive statistics (see Figure S1 for the spatial distributions of prescriptions and green space).

Table 1: Descriptive statistics

| Variables | Total Median | IQR | Rural areas Median | IQR | Urban areas Median | IQR | Wilcoxon $p$-val. |
|---|---|---|---|---|---|---|---|
| Antidep. | 330.50 | 295.00, 361.25 | 341.00 | 309.00, 368.00 | 288.00 | 266.00, 322.00 | <0.001 |
| Green | 75.14 | 57.73, 84.19 | 79.40 | 64.92, 85.59 | 62.67 | 39.56, 75.52 | <0.001 |
| Active | 85.01 | 77.83, 92.04 | 84.98 | 77.91, 92.33 | 85.05 | 77.48, 91.28 | 0.688 |
| Density | 2.04 | 1.10, 4.78 | 1.71 | 0.97, 3.82 | 3.93 | 1.80, 9.63 | <0.001 |
| Elderly | 18.88 | 17.11, 20.71 | 18.92 | 17.28, 20.82 | 18.70 | 16.41, 20.49 | 0.130 |
| GP | 5.90 | 3.20, 11.53 | 4.40 | 2.90, 9.80 | 9.80 | 5.25, 16.70 | <0.001 |
| Housing | 224.00 | 190.00, 257.00 | 215.00 | 185.00, 253.00 | 236.00 | 216.00, 267.50 | <0.001 |
| Non-West. | 3.99 | 2.34, 7.59 | 3.40 | 2.14, 6.60 | 6.10 | 3.59, 12.71 | <0.001 |
| SMR | 1.03 | 0.88, 1.15 | 1.05 | 0.90, 1.16 | 0.98 | 0.85, 1.09 | 0.004 |
| Unempl. | 6.30 | 5.70, 7.03 | 6.30 | 5.80, 7.10 | 6.10 | 5.70, 6.80 | 0.328 |

IQR=interquartile range. *p*-values are based on the Wilcoxon test.

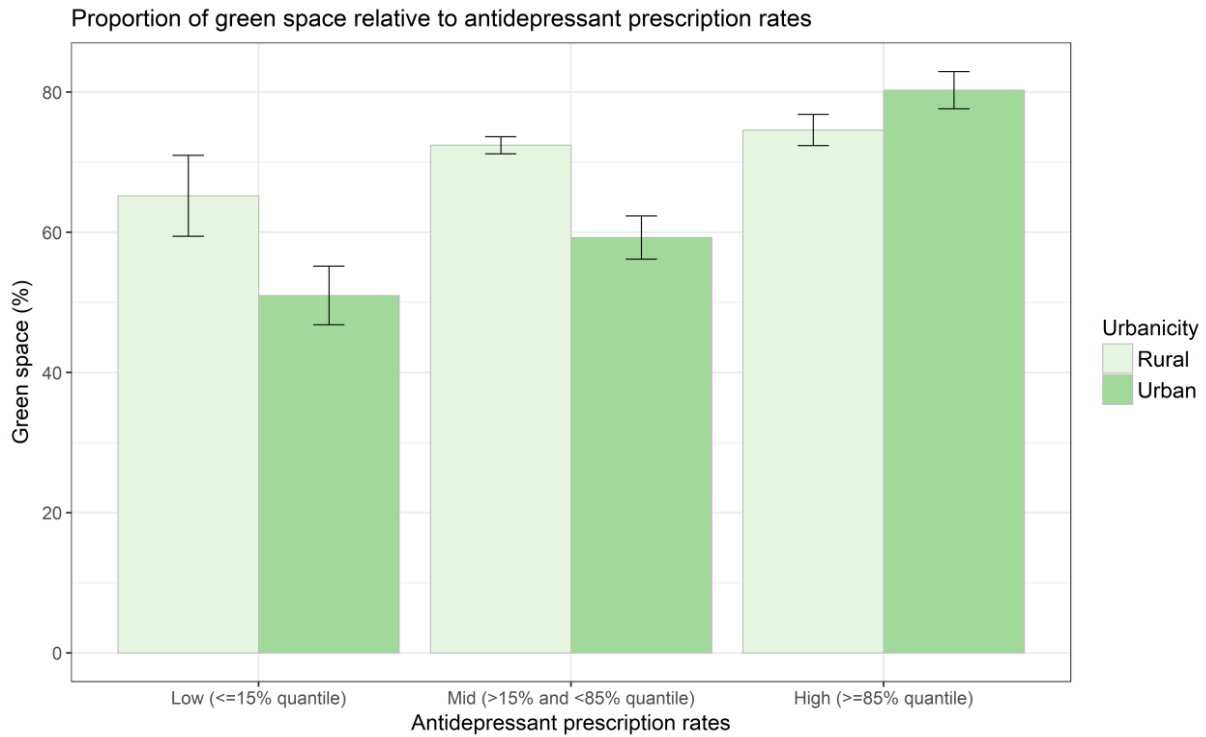

Figure 1: Green space per quantile of antidepressant prescription rates, including the standard errors

Figure 1 summarizes the percentage of green space for different levels of prescription rates subdivided into rural and urban areas. The figure shows that, in general, the more prescriptions, the more green space. Municipalities with low and mid-level prescription rates showed clear differences in the amount of green space. Rural areas with low and mid-level prescription rates have significantly more green space than urban areas. The opposite is observed for areas with high prescription rates.

The correlation analyses indicted that municipalities with higher prescription rates had significantly higher proportions of green space ($\rho=0.262$; $p<0.001$). For the entire correlation matrix, see Table S2 in the supplementary materials. Figure 2 and the bivariate correlations for different prescription levels showed that low prescription rates were insignificantly positively related ($\rho=0.24$, $p=0.067$), mid-level prescription rates were significantly positively related ($\rho=0.15$, $p=0.014$), and high prescription rates were insignificantly negatively related ($\rho=-0.18$, $p=0.174$). The scatterplot smoothers in Figure 2 suggest some non-linearity.

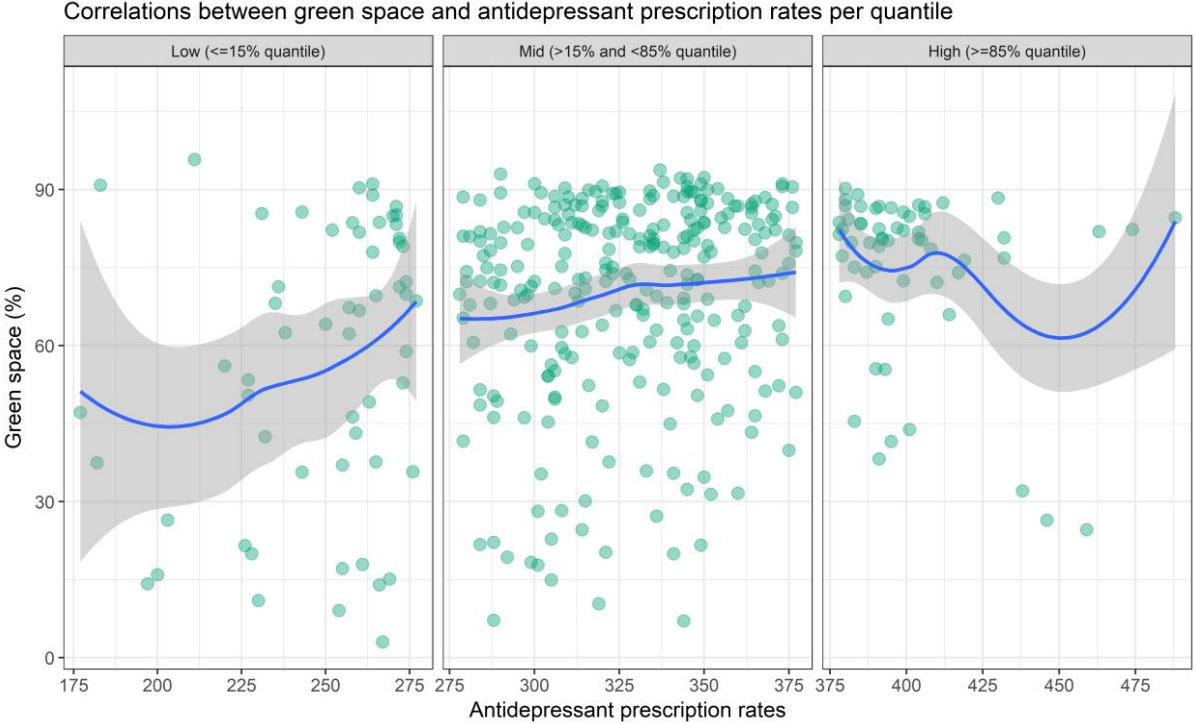

Figure 2: Relations between different levels of antidepressant prescription rates and green space using locally weighted scatterplot smoothers

In order to adjust for confounders, we fitted multivariate BQRs. Table 2 reports the DIC scores of the competing models. Except for the 15% quantile in Model 1, Model 2 is preferred, resulting in lower DIC scores for the 50% and the 85% quantile. Thus, we report the detailed results of Model 2 (see Table S3 for the results of Model 1).

Table 2: BQR model fits (listed are the deviance and DIC values for the chosen quantiles)

|  | Deviance | DIC |
|---|---|---|
| Model 1: 15% quantile | 228.390 | 901.147 |
| Model 2: 15% quantile | 456.199 | 949.220 |
| Model 1: 50% quantile | 347.243 | 1425.090 |
| Model 2: 50% quantile | 329.899 | 1424.080 |
| Model 1: 85% quantile | 695.984 | 825.189 |
| Model 2: 85% quantile | 745.833 | 796.960 |

Figure 3 shows the posterior mean estimates for green space across the 15%, 50%, and 85% quantiles. Moderate deviation in their shapes is noticeable when comparing the three quantiles. However, the green space effect shows a striking significant non-linear pattern. For areas with a low amount of green space, (i.e., below 28%), we found a positive association with antidepressant prescriptions, although the CIs are wide due to only a few observations in this value range. Beyond these values, the green space correlation turned out to have the expected negative association before leveling off. A strong negative effect appears for areas with a large proportion of green space (i.e., >79%).

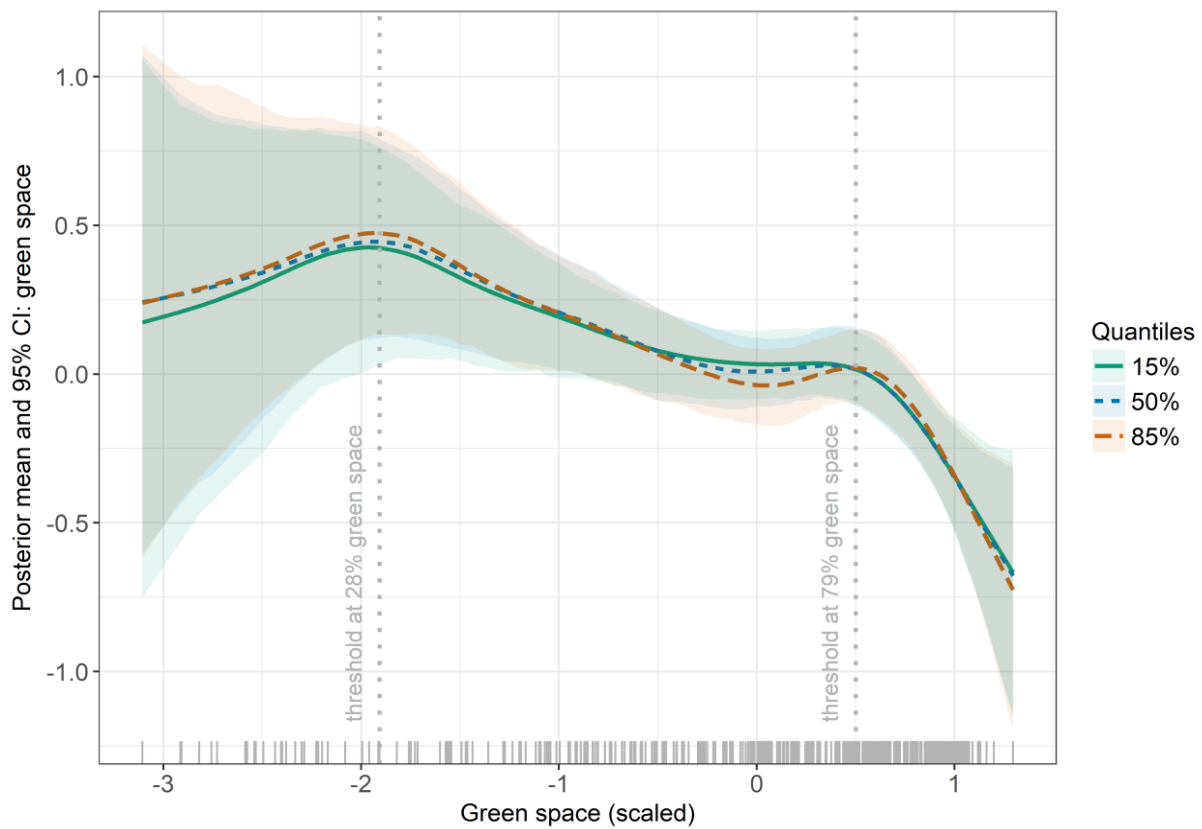

Figure 3: Posterior mean estimates of the non-linear green space effects for different quantiles with the 95% CIs indicated by the shaded areas

Individual covariate associations are given in Figure 4. Physical activity was negatively correlated with prescription rate across the quantiles, though significance was not achieved. Across all quantiles, address density did not reach significance, nor did the GP accessibility. A noticeable difference was apparent for the elderly: This variable was positively related for the 15% and 50% quantiles, but not for the 85% quantile. Areas with higher average housing values were significantly inversely correlated with prescription rates. A large proportion of non-Western people was found to have a pronounced negative association with the antidepressant prescription rate, whereas the posterior mean estimates show some variation from the 15% to the 85% quantile. In contrast, areas with a high standardized mortality ratio and a high unemployment rate were positively correlated with antidepressant prescription rates. Their posterior means are stable across the quantiles. As the covariates were standardized, the model additionally suggested that both area-based unemployment rates and the proportion of non-Western people had a more substantial influence than the standardized mortality ratio and average housing values.

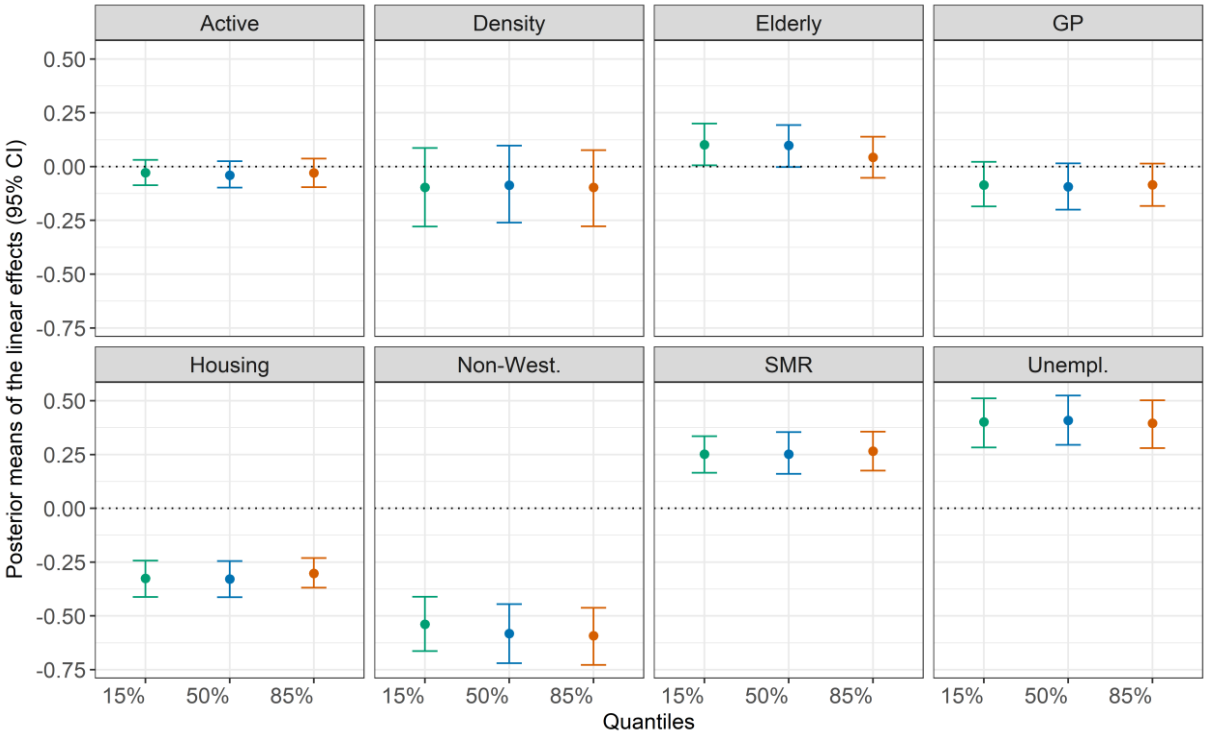

Figure 4: Posterior mean estimates of the covariates in $x$ for the different quantiles together with the 95% CIs

Finally, Figure 5 shows the spatial heterogeneity not explained by the covariates. Red-shaded areas indicate a significant positive spatial effect, while those shaded in blue indicate a significant negative spatial effect. In those areas showing a significant spatial effect, the incorporated variables are less well-suited to explain antidepressant prescription rates.

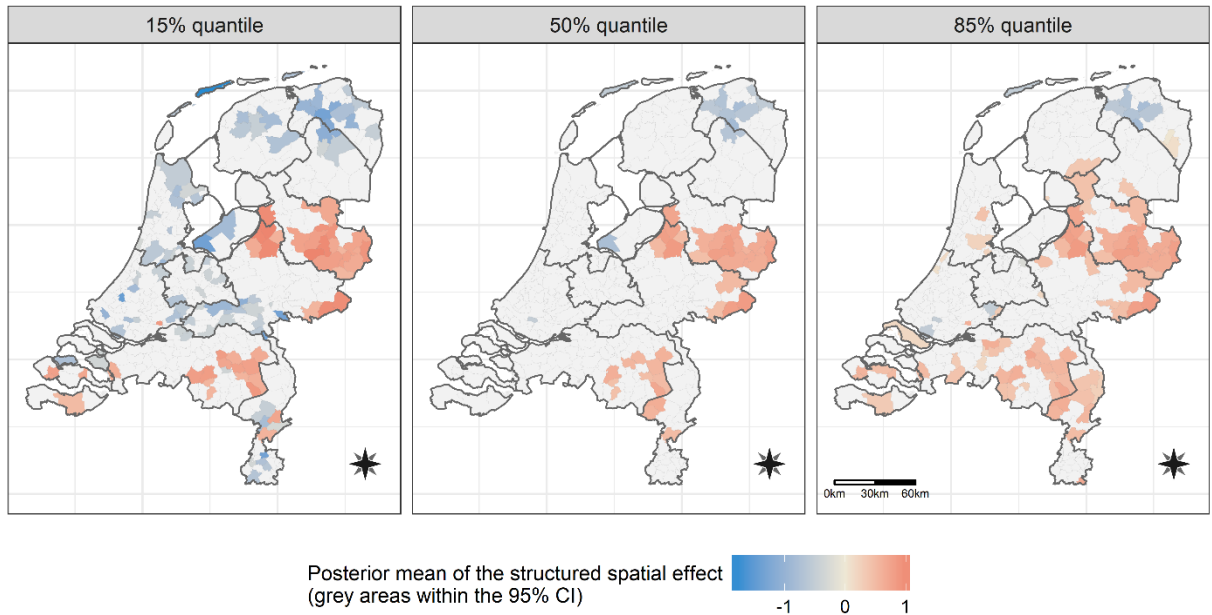

Figure 5: Posterior mean estimates of the structured spatial effect (shown are only the significant effects, i.e., those municipalities for which the estimated posterior 95% CI does not contain the zero. The significant positive effects are shown in red, the negative ones in blue)

## 4 Discussion

### 4.1 Principle findings

This study provided several new insights into green space and mental healthcare consumption in the Netherlands. First, existing knowledge [19,22,28,61–64] was extended by investigating an objective healthcare indicator, namely antidepressant prescription rates, and how these rates were related to exposure to green space. As anticipated, we found a significant negative association between the green space per municipality and the antidepressant prescription rate. In line with a small number of previous studies carried out in the U.K. [35,36,61] and New Zealand [17], ours called attention to Central Europe. While an ecological study in London [35] found similar, though weakly significant, protective effects by correlating greenery (i.e., tree density) with antidepressant prescription rates, our results contrast with a British nationwide analysis that found no such relationship [36]. A study in Auckland City, New Zealand, reported a reduced number of anxiety and mood disorder treatment counts due to a pronounced exposure to green space [17]. Our findings that green space exposures are protective against poor mental health are also corroborated by the mental health literature, as summarized in several reviews and meta-analyses [3,4,26], in general, and by Dutch studies in particular [19,65]. The suggested mechanisms explaining how natural environments such as green space affect mental health are aligned [13,14]. Protective green space effects may operate through, but are not limited to, attention restoration, stress recovery, physical activity, and social interaction [13,14,26,27].

Second, by modeling prescription rate–green space associations through splines using a BQR, we showed significant non-linearities. This finding implies that previously reported relations, assuming a linear association, may not capture the actual effect precisely. Forcing linearity within models, as is typically done [22,28,36,62,65], partly explains the inconsistent findings across studies. Splines [57] are more appealing than a polynomial approach, which requires a priori knowledge of the functional form or the grouping of green space into a few categories, as is frequently done [40]. Our approach is favored as no class breaks need to be set. There have been very few studies specifically confirming non-linear associations [41], which impedes cross-comparisons. Although the differences in the green space–prescription rate associations were moderate across the explored quantiles, so were the differences in the risk and protective factors, BQRs were not restricted in modeling only the mean of the response but the entire distribution [39]. Such models are critical for spatially targeting interventions or tailoring to population groups [62,66].

Third, a striking finding was a dose-response relation between green space and prescription rates. Our results suggest that there are critical values at which green space benefits operate, although this matter is rarely addressed [21,40]. The underlying reasons are still unclear and require further research. Contrasting our initial hypothesis, for green space proportions below 28% we found a positive relation based on wide CIs with antidepressant rates. A plausible explanation is a lack of observations within this data range, as confirmed through the rug plot (Figure 3). Alternatively, despite adjusting for urbanicity – a factor that contributes to differences in the prevalence of mental disorders [56] – it could be that limited green space is caused by competing, possibly unhealthy land use (e.g., more transportation infrastructure causing increased emissions), which negates the health benefits of green [21]. Municipalities with more green space showed the assumed inverse association, corroborating previous studies [3,26]. The findings suggest that the greatest mental health gains may be realized in areas with a proportion of green space of over 79%. Therefore, efforts to increase the quantity of green space available in municipalities should be prioritized by health policymakers.

### 4.2 Strengths and limitations

This study has a number of strengths. To the best of our knowledge, it is the first to propose a BQR model for green space research. While mean regressions are widely applied [28], our model was grounded on the complete distribution of the response, and provided a comprehensive picture of how green space affects individual quantiles of antidepressant prescription rates. This paper also adds to the literature because we relaxed the problematic assumption of a linear association, something that was disregarded in most previous studies [3]. A key strength of our model was the careful consideration of spatial effects, namely that adjacent municipalities might be correlated [35]. As geographic health data become more available on a detailed scale [31], the proposed methodology holds promise for spatial epidemiology. Our study utilized objective healthcare data. As the Netherlands has compulsory health insurance coverage, the data are of high quality. Furthermore, investigating prescription rates at an area level is less sensitive than considering individual GPs, who may have different prescribing habits. Although we tested only one green space specification [36], our results are robust as we extracted the green space information from a high-quality national dataset with a high quality and a coherent methodology.

Notwithstanding these strengths, the present study has some limitations. It needs to be acknowledged that using a cross-sectional design, as is the case for the majority of studies [3,9], prevented us from drawing conclusions about cause–effect relations. Our findings are thus vulnerable to reverse causality, highlighting the need for longitudinal research in future. As most remote sensing-based land use data, ours is limited through a minimum mapping unit [48]. Detailed-scaled green space (e.g., small private gardens) might thus not be captured. Aggregated data per municipality did not allow us to consider heterogeneity in green space availability within a municipality. To address this limitation, we encourage studies on a finer scale, or ideally at an individual level, and recommend considering exposures not only along people's daily mobility, but also over their residential life course [2,67]. As area-level data were used, conclusions cannot be made at the individual level, and the findings might be sensitive to the scale and the aggregation level. Despite having objective data on prescription rates, we were unable to investigate whether people effectively consulted a GP within their municipality, nor can we rule out that people did not take the prescribed antidepressants. This may lead to deviations in the true prescription rates. Of similar importance, prescription data only cover those people who seek help; depressive symptoms often remain unrecognized by non-psychiatric physicians [68]. Finally, although we took steps to collect major confounders, we cannot preclude that our models remain unadjusted for other unmeasured variables.

## 5 Conclusions

This study provided a more nuanced view on the association between antidepressant prescription rates and green space in the Netherlands. Controlling for numerous factors, our findings suggest that more green space is negatively associated with antidepressant prescription rates. This was the first study of its kind showing through BQR that this inverse prescription rate–green space association varies moderately across a lower, a central, and an upper quantile of the response distribution, and the quantiles depend on slightly different control variables. We also took a step forward in demonstrating that the prescription rate–green space association is significantly non-linear across the quantiles, revealing a dose-dependent relation. It seems that there is a threshold of 28% at which green space provides mental health benefits, and most health gains occur when the proportion of green space exceeds 79%. These findings have important implications for planners.

**Ethics statement**

**Acknowledgements**

The NEEDS project has received funding from the European Research Council (ERC) under the European Union's Horizon 2020 research and innovation program (grant agreement No 714993). The funders had no role concerning the study design, data collection and analysis, interpretation, or dissemination. Further details about the study can be found on the NEEDS web-page (http://needs.sites.uu.nl/).

**Author contributions**

MH developed the research idea and study design and collected the data. MH and NK carried out the statistical analysis. With contributions from HR, PH, NK, and PG, MH did the interpretation. MH drafted the manuscript. All authors read and approved the final manuscript.

**Declaration of interest**

None.

**Conflict of interest**

The authors have no conflict of interest to declare.

# Supplementary Materials

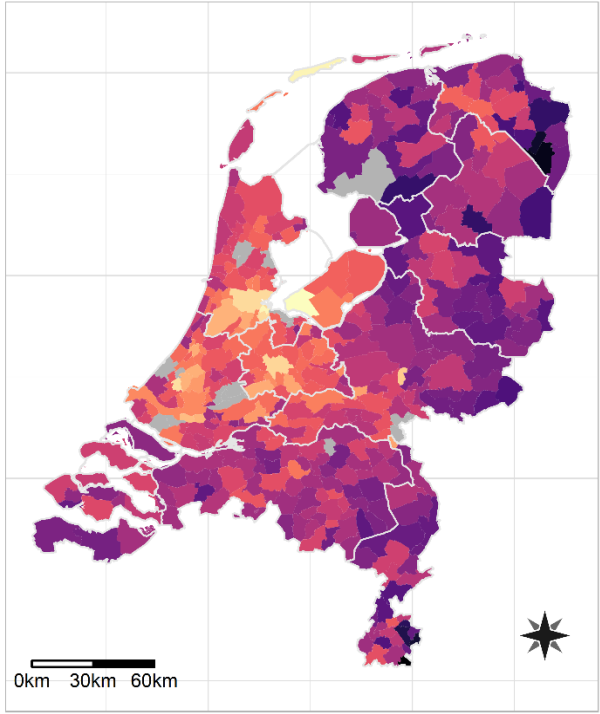
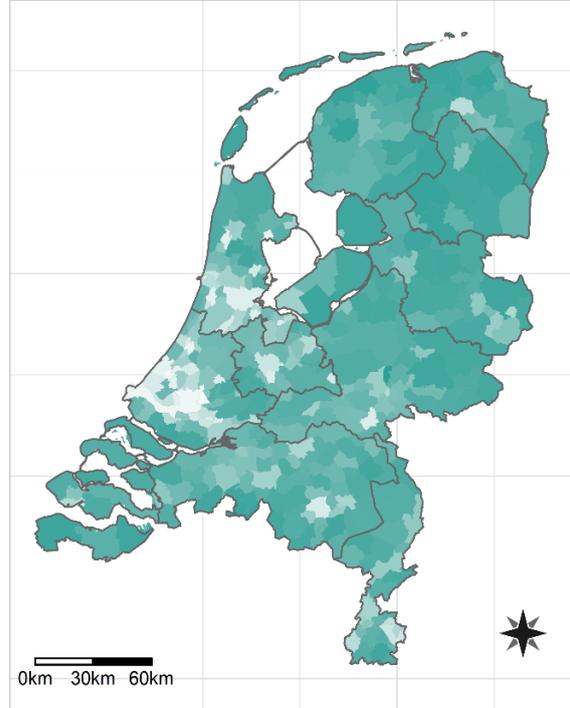
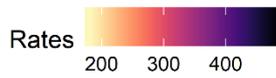
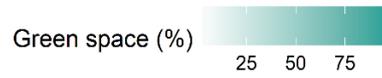

Figure S1: Spatial distribution of antidepressant prescriptions and green space

Table S2: Results of the correlation analyses (*p*-values are given in the lower diagonal)

|            | Antidep. | Green | Elderly | Unempl. | Active | Housing | GP     | Density | Non-West. | SMR    |
|------------|----------|-------|---------|---------|--------|---------|--------|---------|-----------|--------|
| Antidep.   |          | 0.263 | 0.347   | 0.191   | 0.002  | -0.385  | -0.349 | -0.287  | -0.371    | 0.460  |
| Green      | 0.000    |       | 0.250   | -0.392  | -0.012 | 0.128   | -0.720 | -0.908  | -0.707    | 0.069  |
| Elderly    | 0.000    | 0.000 |         | -0.120  | 0.114  | 0.222   | -0.186 | -0.245  | -0.358    | 0.708  |
| Unempl.    | 0.000    | 0.000 | 0.019   |         | 0.073  | -0.531  | 0.423  | 0.413   | 0.628     | 0.192  |
| Active     | 0.963    | 0.812 | 0.025   | 0.152   |        | -0.056  | 0.087  | 0.024   | 0.036     | 0.053  |
| Housing    | 0.000    | 0.012 | 0.000   | 0.000   | 0.275  |         | -0.017 | -0.154  | -0.196    | -0.022 |
| GP         | 0.000    | 0.000 | 0.000   | 0.000   | 0.088  | 0.737   |        | 0.781   | 0.748     | -0.019 |
| Density    | 0.000    | 0.000 | 0.000   | 0.000   | 0.635  | 0.002   | 0.000  |         | 0.730     | -0.042 |
| Non-west.  | 0.000    | 0.000 | 0.000   | 0.000   | 0.478  | 0.000   | 0.000  | 0.000   |           | -0.107 |
| SMR        | 0.000    | 0.176 | 0.000   | 0.000   | 0.297  | 0.672   | 0.704  | 0.412   | 0.036     |        |

The subsequent lines indicate how we estimated the Bayesian geoadditive quantile regression (i.e., Model 2 for the 15%, 50%, and 85% quantiles) using BayesX software.

```
logopen using D:/workspace/log_M2analysis.txt

dataset d
d.infile using D:/workspace/data/dataNL.raw

map m
m.infile, graph using D:/workspace/data/graphNL.gra

mcmcreg yreg
yreg.outfile = D:/workspace/results/M2/q15

yreg.hregress adep = const + green(pspline) + elderly + unemp + acttrans
+ house + gp + dens + nonw + SMR + region(spatial,map=m), hlevel=1
setseed=58581 predict=light family=quantreg quantile = 0.15   using d

yreg.getsample
drop y
drop m
drop yreg

map m
m.infile, graph using D:/workspace/data/graphNL.gra

mcmcreg yreg
yreg.outfile = D:/workspace/results/M2/q50

yreg.hregress adep = const + green(pspline) + elderly + unemp + acttrans
+ house + gp + dens + nonw + SMR + region(spatial,map=m), hlevel=1
setseed=58581 predict=light family=quantreg quantile = 0.5   using d

yreg.getsample
drop y
drop m
drop yreg

map m
m.infile, graph using D:/workspace/data/graphNL.gra

mcmcreg yreg
yreg.outfile = D:/workspace/results/M2/q85

yreg.hregress adep = const + green(pspline) + elderly + unemp + acttrans
+ house + gp + dens + nonw + SMR + region(spatial,map=m), hlevel=1
setseed=58581 predict=light family=quantreg quantile = 0.85   using d

yreg.getsample
drop y
drop m
drop yreg
logclose
```

Table S3: Results of Model 1

| 15% quantile | Posterior mean | 95% CI | |
|---|---|---|---|
| Intercept | -0.235 | -0.378 | -0.026 |
| Green | -0.241 | -0.38 | -0.087 |
| Elderly | 0.086 | -0.021 | 0.181 |
| Unempl. | 0.422 | 0.311 | 0.534 |
| Active | -0.027 | -0.094 | 0.031 |
| Housing | -0.326 | -0.415 | -0.248 |
| GP | -0.073 | -0.172 | 0.031 |
| Density | -0.183 | -0.328 | -0.032 |
| Non-West. | -0.567 | -0.704 | -0.433 |
| SMR | 0.266 | 0.173 | 0.353 |
| 50% quantile | Posterior mean | 95% CI | |
| Intercept | 0.001 | -0.045 | 0.046 |
| Green | -0.243 | -0.387 | -0.093 |
| Elderly | 0.084 | -0.020 | 0.184 |
| Unempl. | 0.420 | 0.296 | 0.529 |
| Active | -0.044 | -0.101 | 0.017 |
| Housing | -0.335 | -0.421 | -0.255 |
| GP | -0.087 | -0.195 | 0.025 |
| Density | -0.164 | -0.321 | -0.013 |
| Non-West. | -0.598 | -0.723 | -0.460 |
| SMR | 0.279 | 0.189 | 0.375 |
| 85% quantile | Posterior mean | 95% CI | |
| Intercept | 0.313 | 0.216 | 0.413 |
| Green | -0.233 | -0.383 | -0.090 |
| Elderly | 0.030 | -0.069 | 0.123 |
| Unempl. | 0.404 | 0.288 | 0.518 |
| Active | -0.032 | -0.092 | 0.031 |
| Housing | -0.312 | -0.389 | -0.241 |
| GP | -0.084 | -0.180 | 0.018 |
| Density | -0.145 | -0.314 | 0.012 |
| Non-West. | -0.611 | -0.753 | -0.478 |
| SMR | 0.290 | 0.201 | 0.377 |